\documentclass[12pt]{article}
\usepackage{graphicx}

\newcommand\pubnumber{DPF2013-122}
\newcommand\pubdate{\today}

\def\stanford{Department of Physics\\
Stanford University, Stanford, CA, U.S.A.}
\def\support{\footnote{On behalf of the \babar\ Collaboration}} 

\def\Title#1{\begin{center} {\Large #1 } \end{center}}
\def\Author#1{\begin{center}{ \sc #1} \end{center}}
\def\Address#1{\begin{center}{ \it #1} \end{center}}

\newcommand\pubblock{\rightline{\begin{tabular}{l} \pubnumber\\
         \pubdate  \end{tabular}}}
\newenvironment{Abstract}{\begin{quotation}  }{\end{quotation}}
\newenvironment{Presented}{\begin{quotation} \begin{center} 
             PRESENTED AT\end{center}\bigskip 
      \begin{center}\begin{large}}{\end{large}\end{center} \end{quotation}}
\def\Acknowledgments{\bigskip  \bigskip \begin{center} \begin{large}
             \bf ACKNOWLEDGMENTS \end{large}\end{center}}

\def\beq{\begin{equation}}
\def\eeq#1{\label{#1}\end{equation}}
\def\eeqn{\end{equation}}

\def\beqa{\begin{eqnarray}}
\def\eeqa#1{\label{#1}\end{eqnarray}}
\def\eeqan{\end{eqnarray}}

\let\bar=\overbar

\def\Dslash{\not{\hbox{\kern-4pt $D$}}}
\def\dslash{\not{\hbox{\kern-2pt $\del$}}}

\def\BR{\mbox{\rm BR}}

\def\msb{{\bar{\ssstyle M \kern -1pt S}}}

\def\s#1{\widetilde{#1}}

\input babarsym.tex
\graphicspath{{figures/}}

\newcommand{\eqref}[1]{Eq.~(\ref{eq:#1})}

\newcommand{\rhom}               {\mbox{$\rho^-$}}

\begin{document}
\begin{titlepage}
\pubblock

\vfill
\Title{Lepton-number violation in \B\ decays at \babar}
\vfill
\Author{ Eugenia Maria Teresa Irene Puccio\support}
\Address{\stanford}
\vfill
\begin{Abstract}
We present results of searches for lepton-number and, in some cases also
baryon-number, violation in \B\ decays using the full \babar\ dataset of $471$
million \BB\ pairs. 
\end{Abstract}
\vfill
\begin{Presented}
DPF 2013\\
The Meeting of the American Physical Society\\
Division of Particles and Fields\\
Santa Cruz, California, August 13--17, 2013\\
\end{Presented}
\vfill
\end{titlepage}
\def\thefootnote{\fnsymbol{footnote}}
\setcounter{footnote}{0}

\section{Introduction}

\begin{figure}[htb]
\centering
\includegraphics[height=1.8in]{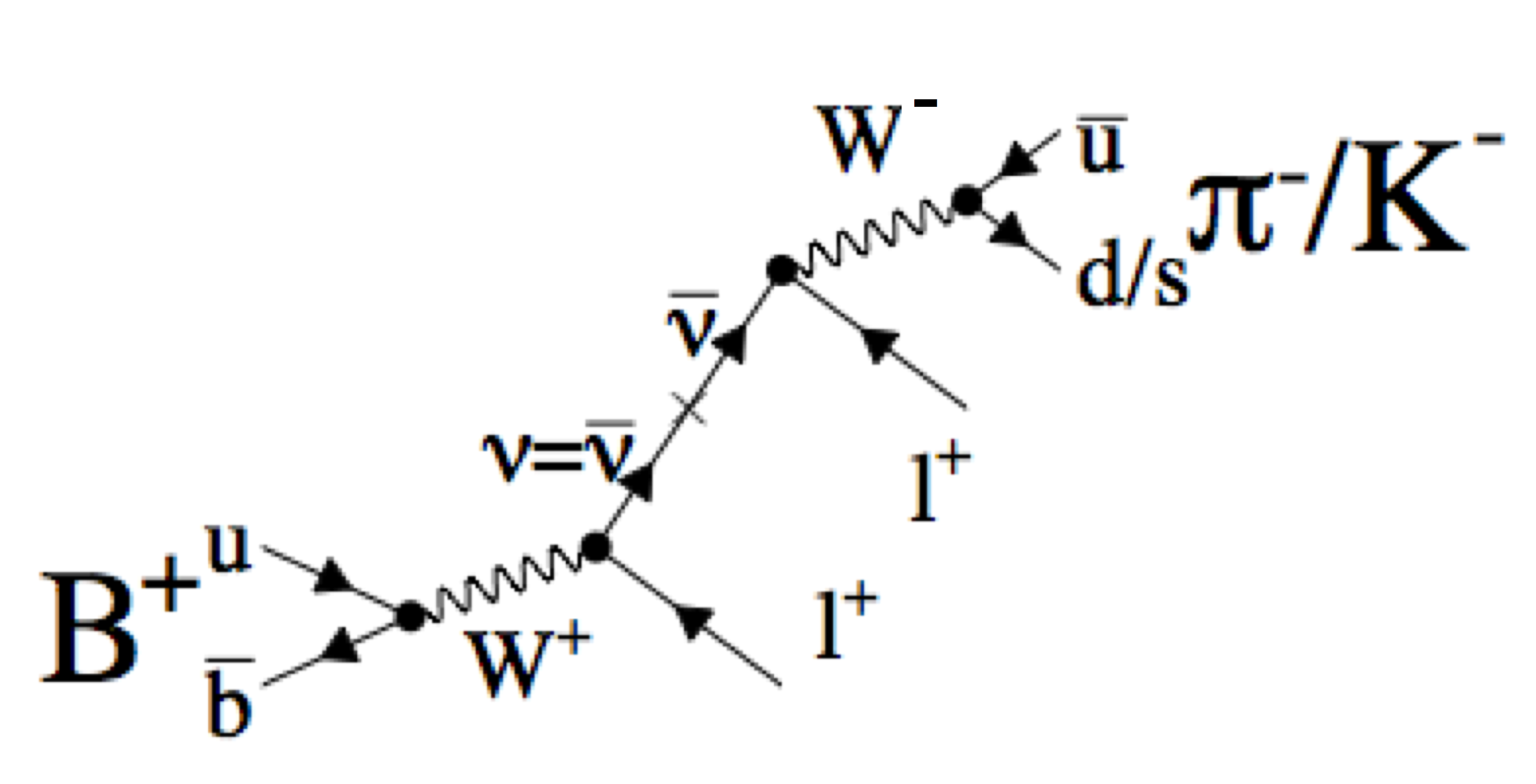}
\caption{Example diagram of a process with $\Delta L=2$ due to the exchange
of a Majorana neutrino\cite{BABAR:2012aa}.}
\label{fig:feynmann-example}
\end{figure}

The observation of neutrino oscillations suggests that lepton flavour is
not a conserved quantity. If neutrinos have mass, then the neutrino and
antineutrino can be the same particle and processes involving lepton-number
violation become possible. The most sensitive
searches for lepton-number violation are currently found in neutrinoless double-beta
decays~\cite{GomezCadenas:2011it}. However the nuclear environment of this
type of search makes it difficult to extract the neutrino mass scale.
Processes involving mesons decaying to lepton-number violating final states
have been suggested as an alternative to the neutrinoless double-beta decay
searches. \figref{feynmann-example} shows an example Feynman diagram for a
\B\ decay to a lepton-number violating final state via the exchange of an
s-channel Majorana neutrino. It is however expected that these decays have
extremely small unobservable probabilities and therefore the observation
of a significant signal would be a clear sign of New Physics. 

In this paper, we present the results of analyses conducted by the \babar\
collaboration for the following lepton-number violating \B\ decays:
$\Bz\to\Lambda_{c}^{+}\ell^{-}$, $\Bp\to\Lambda^{0}\ell^{+}$ and $\Bp\to
h^{-}\ell^{+}\ell'^{+}$, where $\ell^{+}$ and $\ell'^{+}$ are either \ep\
or \mup\ leptons and $h$ is either a kaon, pion, \Kstar, $\rho$, or $D$
meson.  

These analyses make use of the full \babar\ \FourS\ dataset. This consists
of $471$ million \BB\ pairs collected by the \babar\ detector at the \pep2\
\BF, which collides \epem\ asymmetric-energy beams at the \FourS\
resonance~\cite{TheBABAR:2013jta}. The \B\ meson candidates are
characterised by two kinematic variables. We take advantage of the
precise kinematic information for the beams to form the variables
$\mes=\sqrt{\frac{s}{4}-p^{*2}_{\B}}$ and
$\DeltaE=E_{\B}^{*}-\frac{\sqrt{s}}{2}$, where
$\left(E^{*}_{\B},\vec{p}^{\,*}_{\B}\right)$ is the \B\ meson four-momentum
in the \epem\ centre-of-mass frame and $\sqrt{s}$ is the centre-of-mass
energy.  Signal events are expected to peak around the \B\ mass for \mes\
and around zero for \DeltaE. There are two main types of backgrounds: the
very abundant \qqbar\ events, where \q\ is either a \u, \d, \s\ or \c\
quark, and the background arising from \B\ decays to other final states. To
distinguish \B\ meson candidates from the continuum \qqbar\ background,
variables describing the topology of the event are combined in a
multivariate analyser (MVA), such as a neural network or a Fisher
discriminant, in order to maximise their discriminating power.  The
variables \mes, \DeltaE\ and the output of the MVA either have selection
requirements placed upon them or are supplied as inputs to a maximum
likelihood fit. 
 
\section{Lepton and baryon number violation in \\
$B\to\Lambda_{(c)}\en(\mun)$ decays}

\begin{figure}[htb]
\centering
\includegraphics[height=2in]{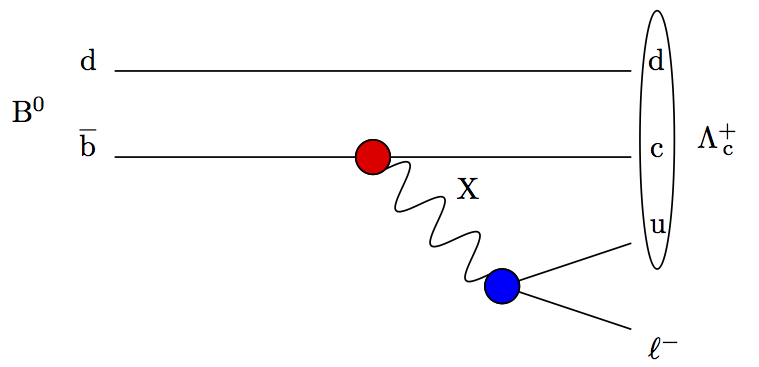}
\caption{Feynman diagram for the $\Bz\to\Lambda_{c}^{+}l^{-}$ decay.}
\label{fig:lambdac-feynman}
\end{figure}

This is the first and only search for the decay modes
$\Bz\to\Lambda_{c}^{+}\ell^{-}$ and $\Bm\to\Lambda^{0}\ell^{-}$, where $\ell^{-}$
is either an electron or a muon. These
decays are particularly interesting since a significant signal would not
only be an indication of lepton-number violation but also of baryon-number
violation. \figref{lambdac-feynman} shows a typical Feynman diagram for the
$\Bz\to\Lambda_{c}^{+}l^{-}$ decays. We reconstruct $\Lambda_{c}^{+}\to
p\Km\pip$, which has a branching fraction of ${\cal B}\approx5\%$, and
$\Lambda^{0}\to p\pim$,
with branching fraction ${\cal B}\approx64\%$. The baryon candidate is
required to have originated from a common vertex with a lepton. A candidate
event is also required to have more than four charged tracks.  A neural
network composed of six event shape variables is used to reduce remaining
backgrounds. 

\begin{figure}[htb]
\centering
\includegraphics[height=4in]{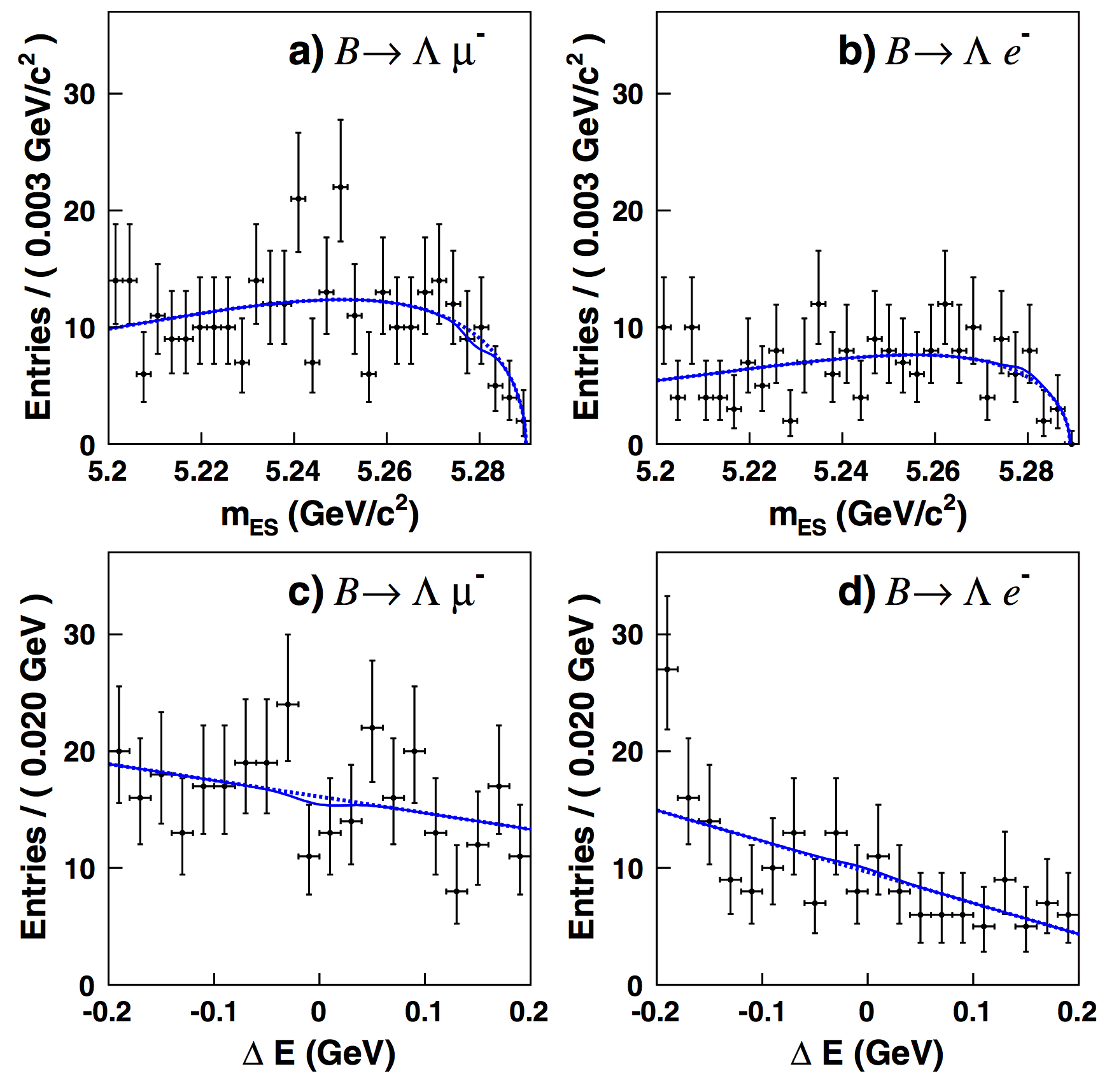}
\caption{\mes\ and \DeltaE\ projection distributions for the decays
	$\Bm\to\Lambda^{0}\ell^{-}$, where $\ell$ is either an electron or
	a muon. The blue solid line is the overall fit and the black points
	with error bars correspond to data.
}
\label{fig:lambda-projections}
\end{figure}

The signal \mes\ and \DeltaE\ distributions are parametrised in the fit to
data with a Crystal Ball function PDF. For the $\Lambda_{c}^{+}$ decay
modes, an additional parametric PDF is added to the fit to parametrise the
distribution of the output of the MVA. \figref{lambda-projections} shows
projection plots for \mes\ and \DeltaE\ for the $\Bm\to\Lambda^{0}\ell^{-}$
decays, where the blue line is the overall fit and the black points with
error bars correspond to the data distribution. We
observe no significant signals in any of the modes and measure branching
fraction upper limits in the range of $(3-180)\times10^{-8}$ at $90\%$
confidence level. The most sensitive decay was found to be
$\Bm\to\Lambda^{0}\en$ with a branching fraction upper limit of ${\cal
B}_{90\%}<3.2\times10^{-8}$~\cite{BABAR:2011ac}.

\section{Lepton-number violation in $\Bp\to h^{-}\ell^{+}\ell'^{+}$}

Two different analyses were performed by \babar, each with a different set
of selection criteria. The first looked for the decays $\Bp\to
h^{-}\ell^{+}\ell^{+}$,
where $h$ is either a pion or a kaon and $\ell$ is a same-flavour electron or
muon. The decay mechanism is fairly similar to the neutrinoless double-beta
decays. Again candidate events are selected with more than four charged tracks
originating from the same vertex and with same-sign charged leptons in the
final state. In this analysis a veto is applied to remove backgrounds
originating from the decay of a \B\ meson to \jpsi\ or \psitwos\ and a pion
or kaon. Other types of backgrounds are reduced by forming a boosted
decision tree using $18$ event-shape variables.

\begin{figure}[htb]
\centering
\includegraphics[height=3in]{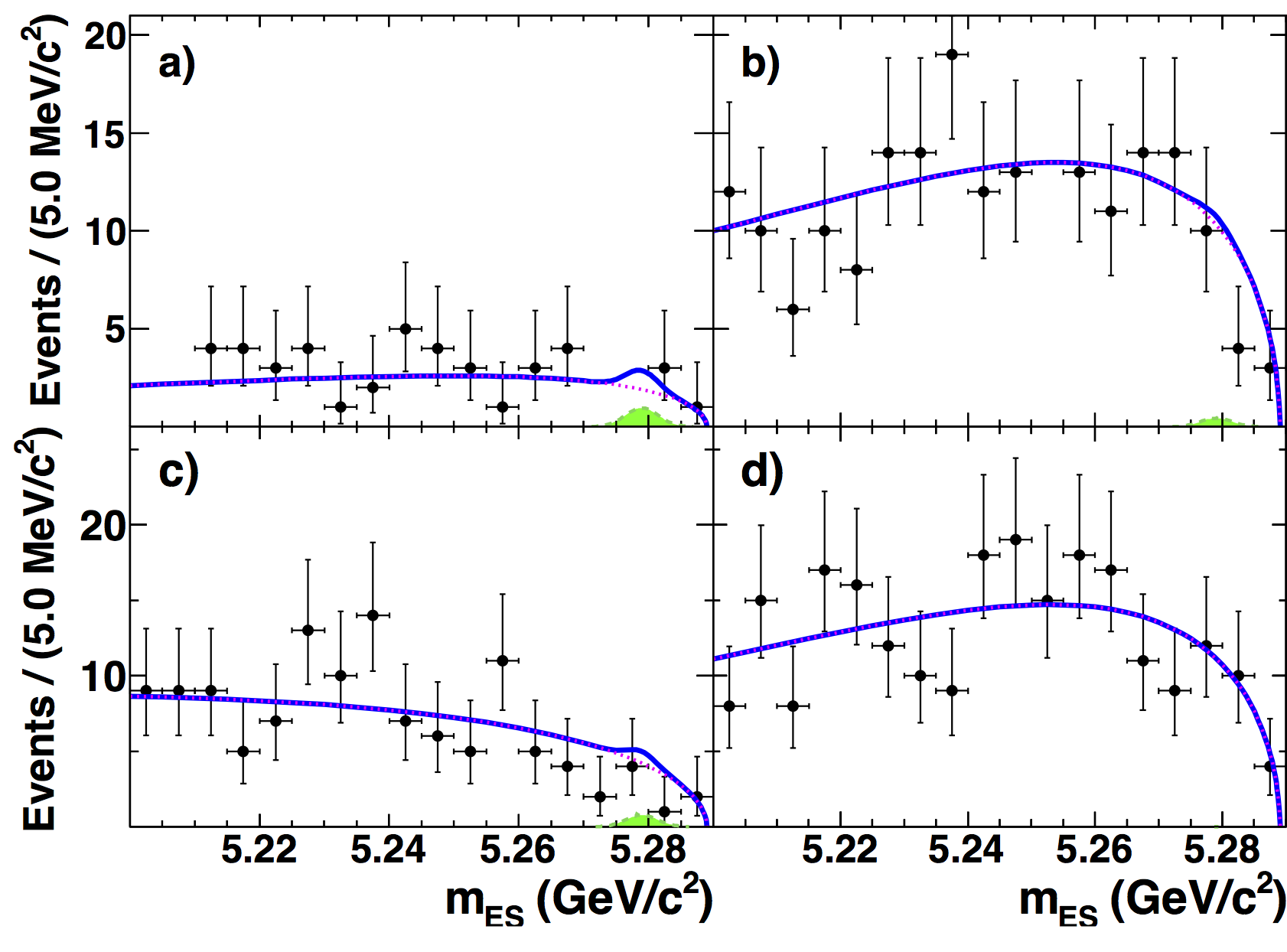}
\caption{\mes\ projection distributions for the decays
	$\Bp\to h^{-}\ell^{+}\ell^{+}$, where $h$ is either a kaon or a pion and
	$\ell$ a same-flavour lepton. The blue solid line is the overall fit,
	the solid green shape the overall signal and the black points with
	error barsthe data.}
\label{fig:korpill-projections}
\end{figure}

\begin{figure}[htb]
\centering
\includegraphics[height=2.5in]{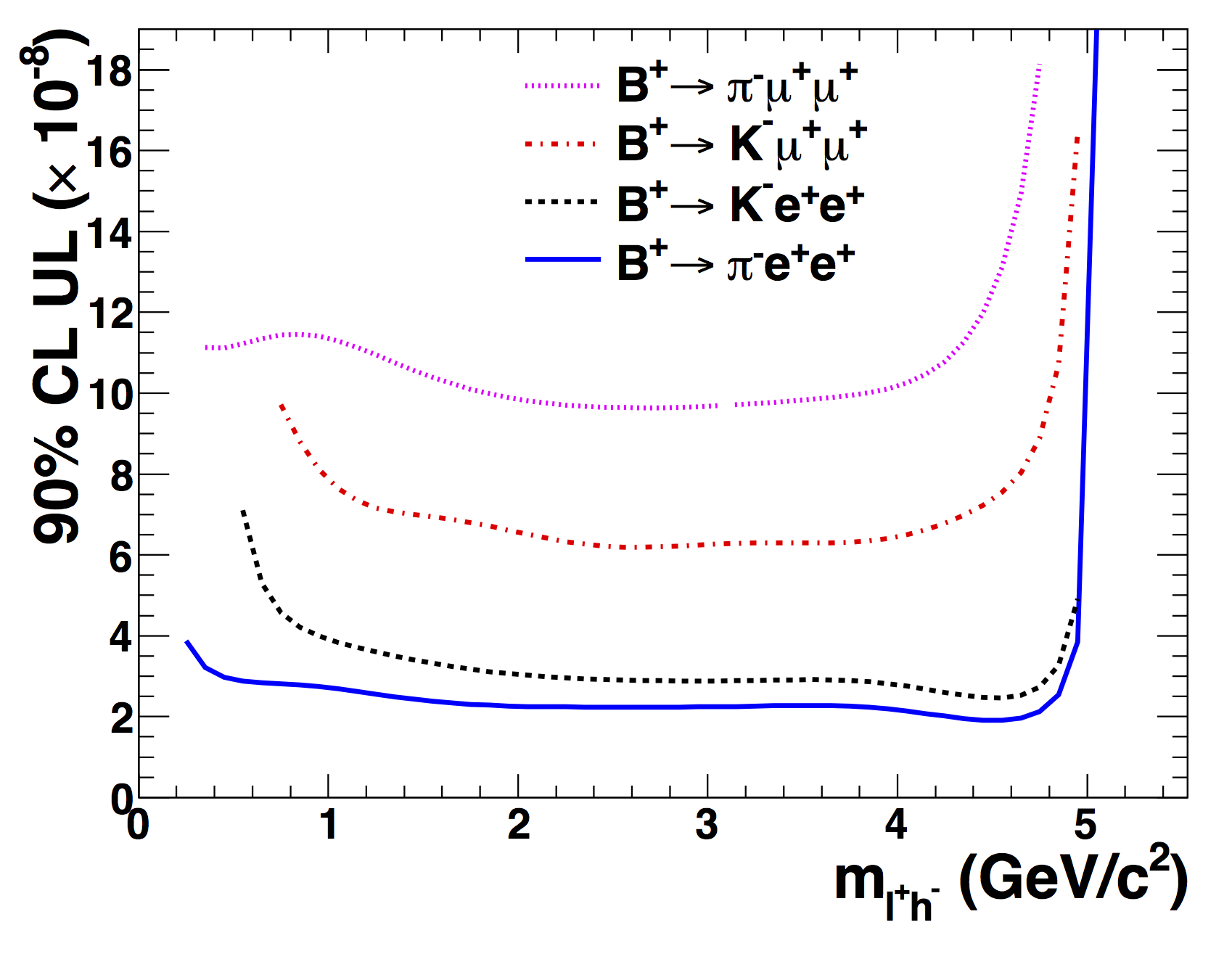}
\caption{Upper limits for the $\Bp\to h^{-}\ell^{+}\ell^{+}$ decays plotted
against the $h^{-}\ell^{+}$ invariant mass. If the decay process proceeds
via the exchange of a Majorana neutrino then this quantity is
related to the neutrino mass.}
\label{fig:upperlimit-mass}
\end{figure}

The signal \mes\ and \DeltaE\ distributions are parametrised using a
Gaussian PDF where the mean and width parameters are determined from a fit
to the kinematically similar decay $\Bp\to\jpsi(\ell^{+}\ell^{-})h^{+}$.
\figref{korpill-projections} shows the resultant fit (blue
solid line) over the data points for \mes\ for all the $\Bp\to
h^{-}\ell^{+}\ell^{+}$ decay modes. The green area is the overall signal. We do not
observe significant signals for any of these modes and find $90\%$
confidence level upper limits of the order of $10^{-8}$. In
\figref{upperlimit-mass}, the branching fraction upper limits are
plotted against the $\ell^{+}h^{-}$ invariant mass. If the decay of $\Bp\to
h^{-}\ell^{+}\ell^{+}$ is governed by the exchange of a Majorana neutrino, then
$m_{\ell^{+}h^{-}}$ is directly related to the neutrino
mass~\cite{BABAR:2012aa}.

The second analysis, performed recently by \babar, searched for 11 decay
modes:
$\Bp\to\rhom \ell^{+}\ell'^{+}$,
$\Bp\to\Kstarm(\to\KS\pim(\Km\piz))\ell^{+}\ell'^{+}$,
$\Bp\to\Dm(\to\Km\pim\pip)\ell^{+}\ell'^{+}$ and $\Bp\to\Km(\pim)\ep\mup$. The
difference here is that the lepton $\ell$ may not be of the same flavour.
These types of decays can occur if the Majorana neutrino oscillates during
the exchange.  Hence $\Bp\to h^{-}\ep\mup$ decays, where $h$ here can be
any of the above listed final-state mesons, are also lepton-flavour
violating. Most of these decays, except for the few listed above, have only
been previously studied by the CLEO Collaboration, which published upper
limits at $90\%$ confidence level for final states containing a $\pi$, $K$, $\rho$ or $\Kstar$
of order $(1.0-8.3)\times10^{-6}$~\cite{Edwards:2002kq}. The Belle
Collaboration, and LHCb in the $\mup\mup$ case only, both measured upper
limits for the decays to $D$ mesons of the order of
$10^{-6}$~\cite{Seon:2011ni}~\cite{Aaij:2012zr}. 

\begin{figure}[htb]
\centering
\includegraphics[height=1.8in]{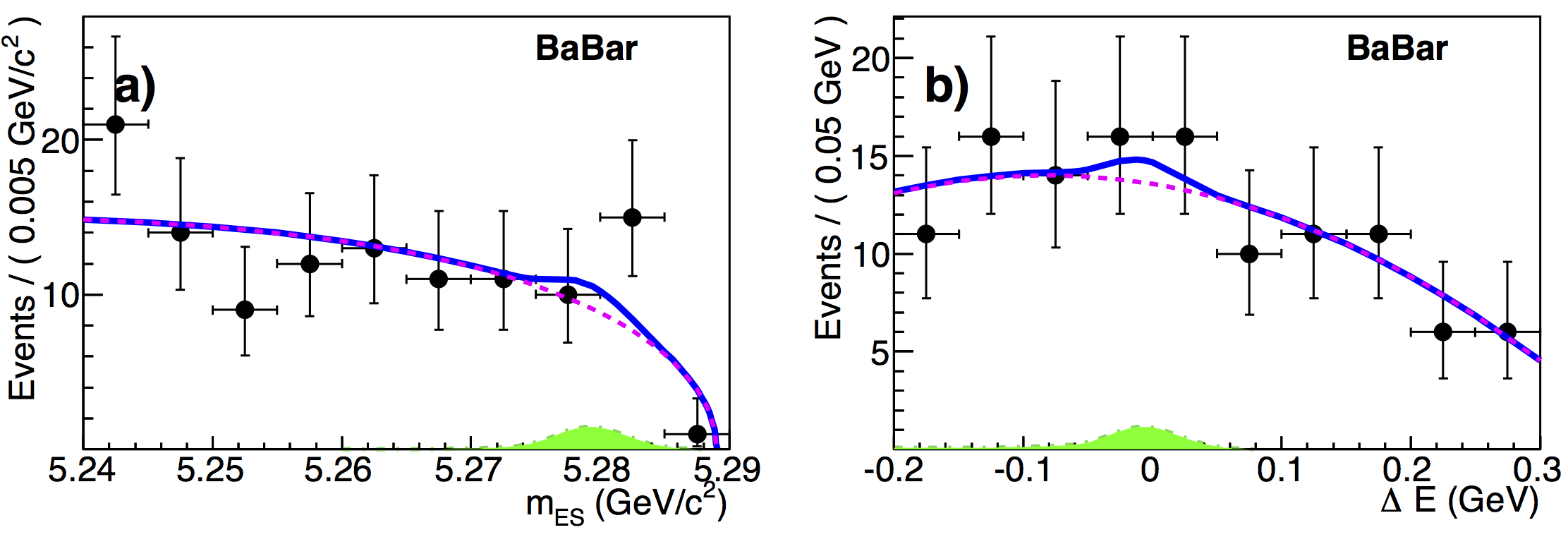}
\caption{\mes\ and \DeltaE\ projection distributions for the decay
	$\Bp\to\Km\ep\mup$. The blue solid line is the overall fit,
	the solid green shape the overall signal and the black markers
	indicate the data points.}
\label{fig:kepmup-projections}
\end{figure}

\begin{figure}[!h]
\centering
\includegraphics[height=2.4in]{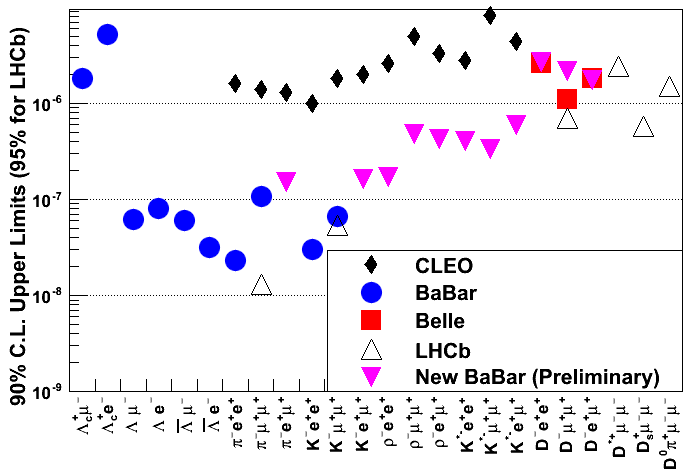}
\caption{Summary of all $90\%$ confidence-level upper limits for all the
lepton-number violating \B\ decays measured. Blue circles indicate the
published \babar\ measurements and solid magenta triangles the new
preliminary results. Other measurements include the
CLEO Collaboration (black diamonds), Belle (red squares) and LHCb (open
triangles). Note that LHCb upper limits are quoted at $95\%$ confidence
level.}
\label{fig:upperlimits}
\end{figure}

The analysis follows a similar procedure to the previously mentioned analysis, only that
selections are tailored to select efficiently all 11 decay modes in one
selection procedure. The boosted decision tree in this analysis only
includes 9 event-shape variables. Signal \mes\ and \DeltaE\
distributions are modelled in the fit to data using a Crystal Ball
function. \figref{kepmup-projections} shows projections of fit variables for
data, for the fit (blue solid line) and overall signal (green solid area) using
$\Bp\to\Km\ep\mup$ as an example. Preliminary results show that no
significant signal is found in any of the channels, and $90\%$ confidence
level upper-limits are set ranging between $(1.5-26.4)\times10^{-7}$.
\figref{upperlimits} shows a plot of all the upper-limit measurements to
date made by the different collaborations, including the new \babar\
preliminary upper limits, which are indicated with solid magenta triangles.
We have the most stringent upper limits for the following decays:
$\Bp\to\pim\ep\mup$, $\Bp\to\Km\ep\mup$, and all the decays to $\rho$ or
$\Kstar$. Our results for the upper limits in the decays to a $D$ meson are
comparable to the results obtained by both Belle and LHCb. This plot also
shows that we still hold the most stringent upper limits in the decays to
$\Lambda_{(c)}$ and $\Km(\pim)\ep\ep$. Our $\Bp\to\Km\mup\mup$ upper limit
is comparable to the upper limit obtained by the LHCb Collaboration.

\section{Conclusion}

\babar\ has searched for many lepton-number violating decays; however no
significant signal was found for any of the decay modes. 
Our measurements give more stringent $90\%$ confidence level
upper limits compared to the other experiments for most of the decays
studied.  Our best sensitivity to branching fractions for these modes is of
the order of $10^{-8}$.  Higher luminosity experiments such as Belle-II and
LHCb can still increase the sensitivity to these decays. However, the
higher background levels can make study of these rare decays more
difficult.  Some future experiments will be specifically designed to look
for lepton-number violation and lepton-flavour violation.  One of these is
the Mu2e experiment at Fermilab, which will search for muons converting to
electrons. The goal of the experiment is to be able to reach a sensitivity
to this type of process of the order of $10^{-17}$.  Other experiments
specifically designed for these types of searches include neutrinoless
double-beta decay experiments. 

\Acknowledgments

I would like to thank Prof. Patricia Burchat, Prof. Brian Meadows, Dr.
Fergus Wilson and Dr. Matt Bellis for all the help and suggestions given to
me in the preparation for this talk and the proceedings.

\end{document}